\documentclass{article}[15pt]

\usepackage[dvips]{epsfig}
\usepackage{rotating}

\renewcommand{\baselinestretch}{1.35} % double spacing

\begin{document}

\title{
An Unusual 500,000 Bases Long Oscillation of Guanine and Cytosine Content
in Human Chromosome 21 
\vspace{0.2in}
\author{
Wentian Li$^{a}$\footnote{Corresponding author. 
Email: wli@nslij-genetics.org.} 
and  Dirk Holste$^{b}$ \\
{\small \sl  a. The Robert S. Boas Center for Genomics and Human Genetics}\\
{\small \sl North Shore LIJ Institute for Medical Research, Manhasset, NY 11030, USA.}\\
{\small \sl b. Department of Biology, Massachusetts Institute of Technology, 
Cambridge, MA 02139, USA.}
}
\date{}
}
\maketitle  % End title section
\markboth{\sl W.Li) }{\sl W.Li }

\large

\begin{center}
{\bf Abstract}
\end{center}

An oscillation with a period of around 500~kb 
in guanine and cytosine content (GC\%) is observed in
the DNA sequence of human chromosome 21. This oscillation
is localized in the rightmost one-eighth region of the chromosome,
from 43.5~Mb to 46.5~Mb.  Five cycles of oscillation  are 
observed in this region with six GC-rich peaks and five GC-poor 
valleys. The GC-poor valleys
comprise regions with low density of CpG islands and,
alternating  between the two DNA strands, low gene density 
regions. Consequently, the long-range oscillation of GC\% 
result in spacing patterns of both CpG island density, and 
to a lesser extent, gene densities.

\newpage
\section{Introduction}

\indent

Periodicities  and characteristic length scales in biological sequences 
have been of long-standing interest in sequence analysis. 
The codon structure and the corresponding length scale of 
three bases in protein coding sequences 
(Shulman et al., 1981; Fickett, 1982; Staden and McLachlan, 1982)
is one of the main features being used for computational 
gene recognization (see, e.g., 
Borodovsky et al., 1986; Borodovsky and McIninch, 1993; Burge and 
Karlin, 1997; Tiwari et al., 1997; Guigo, 1999; Grosse et al., 2000;
Li et al., 2002).
by a tendency for certain base types to be at certain codon 
positions. A correlation at a distance of 10-11 bases was also detected 
Trifonov and Sussman, 1980; Baldi et al., 1996; Widom, 1996;
Tomita et al., 1999). Two possible
explanations have been put forward to explain this (Herzel et al., 1998).
The first explanation is that this periodicity is related to 
DNA bending and nucleosome formation (Trifonov and Sussman, 1980). The
second explanation is that it is a reflection of a periodicity 
in protein sequences (Zhurkin, 1981), because most sequences 
that exhibit 10-11 base correlations are protein coding sequences
(see, however, the result in (Holste et al., 2003), where a 10-11 
base correlation is detected in mostly non-coding human
chromosomes 20,21,22 sequences).
Proponents of the second explanation also argue that 
nucleosome formation is only aided by a sequence 
property on 5-6 bases (Zhurkin, 1981). Periodicities in protein
sequences are seen to be natural, because their presence aids
the secondary structure (Shiba et al., 2002).  Besides these two 
well known length scales in DNA sequences, 3 and 10-11 bases, 
other length scales, e.g., 120, 200, and 450 bases, 
were also proposed, based mainly on a theoretical argument on the 
size of an exon, the size of nucleosome unit, and the size of 
a typical prokaryote protein (Trifonov, 1998). 

Tandem repeats of DNA segments introduce local or even global
sequence periodicities depending on their distribution.
If a sequence of $k$ bases tandemly repeats many times, there
would be base-base correlations at separations of
$k$, $2k$, $3k, \dots$ bases, e.g.,  the periodicity 
of $k=$2 bases observed in non-coding sequences 
(Arques, 1987; Konopka et al., 1987).
If the repeat is not perfect, such as those in many subtelometric 
sequences, a correlation
may not appear at the exact multiples of the basic unit size
(Pizzi et al., 1990). Interspersed repeats in mammalian genomes
(Smit, 1999)  should in principle not introduce characteristic
length scales in correlation patterns because their spatial distribution 
is not regular (except for length scales shorter than
the size of one copy of the repeat). A recent survey of 
characteristic length scales 
in many eukaryote genome sequences by spectral analysis
reveals peaks in power spectra at about 68, 59 and 94 bases in 
{\sl C. elegans} (chromosomes I, II, and III, respectively), 
at about 248, 167, 126 bases in {\sl A. thaliana} chromosome 3, 
at about 174, 88, 59 bases in chromosome 4, 
at about 356, 174, 88, 59 bases in chromosome 5, 
and at about 167, 84 bases in {\sl H. sapiens} chromosomes 21 and 22 
(Fukushima et al., 2002).  A connection between these length scales 
and tandem or interspersed repeats for the three human
chromosomes (20, 21, and 22) has been discussed in (Holste et al., 2003).

At distances much longer than hundreds of bases, it is more
difficult to observe a correlation at the level of
individual bases. It is, however, easier to observe correlations 
between base compositions, such as the guanine and cytosine content 
(GC\%). The reason is as follows: instead of requiring a matching 
base by base at the exact spacing, correlation at the base
composition level only requires GC\% to be similarly 
high (or low) at certain range of spacings. In this paper, we report 
an unusual long-range oscillation of GC\% with a periodicity
of around 500~kb (1~kb= $10^3$ bases) in the DNA sequence of
human chromosome 21.  This periodicity is longer than any 
periodicity in DNA sequences detected so far.

\section{The DNA sequence of human chromosome 21 exhibits
higher correlations}

\indent

Sequence data were downloaded from the UCSC human genome
repository (available at {\sl http://genome.ucsc.edu/}),
for the version of NCBI build-34 release.  We evenly partition 
each human chromosome into $N=2^k$ non-overlapping windows
(e.g., $k=17$ and $N=$131,072). GC\% of each window is 
calculated, forming a GC\% series:
$\{ x_i \}$ ($i=1, 2, \dots N$). The correlation
function $\Gamma(d)$ of this series is defined as the Pearson's
correlation coefficient of two truncated subseries: 
a right-hand side truncated $\{ x' \} \equiv \{ x_i \}$ 
($i=1, 2, \dots N-d$), and a left-hand side truncated 
$\{ x''\} \equiv \{ x_i \}$ ($i=d+1, d+2, \dots N$):
\begin{equation}
\Gamma_w(d) \equiv  \frac{\mbox{Cov}(x', x'')}{\sqrt{ \mbox{Var}(x')} \sqrt{ \mbox{Var}(x'')}}
\end{equation}
where the covariance is defined as Cov$(x',x'') = 
\langle(x'- \langle x'\rangle)(x''- \langle x'' \rangle)\rangle$
and the variance is defined as $\mbox{Var}(x)= \langle (x- \langle x \rangle)^2 \rangle$
($\langle \rangle$ is the average of the $\{ x\}$ series). Here, the parameter
$w$ is used to indicate the fact that the GC\% series, and thus
detected patterns of correlation, implicitly depends on the window 
size $w$.  When the spacing $d \ll N$, the following approximation
formula can be used:
\begin{equation}
\label{eq2}
\mbox{Var}(x') \approx \mbox{Var}(x'') \approx  \mbox{Var}(x),
\hspace{0.5in}
\Gamma_w(d) \approx  \frac{\mbox{Cov}(x', x'')}{\mbox{Var}(x)}
\end{equation}
For each human chromosome, Fig.1 shows $\Gamma(d)$ for both the GC\% 
series derived from $2^{17}$ windows (bottom) and the GC\% series 
derived from $2^{15}$ windows (top). Fig.1 shows 
that the magnitude of $\Gamma(d)$ depends on the window size.

We illustrate how the magnitude of the correlation function 
depends on the window size  by plotting in Fig.2  the correlation 
$\Gamma_w(d)$ at distance close to $d \approx$ 1~Mb 
(1~Mb = $10^6$ bases) for each human chromosome versus the 
chromosome-specific window size $w$.
All $w$ values are within the range of 0.3-2~kb.  In a 
double-logarithmic representation, $\log (\Gamma_w(d=1~\mbox{Mb}) )$
for the majority of human chromosomes follows a linear function
of $\log (w) $, and thus $ \Gamma_w(d=1~\mbox{Mb}) \sim w^b$ ($b > 0$)
approximates a power-law function.

The dependence of $\Gamma_w(d)$ on the window size $w$ shown
in Fig.2 was previously observed (Li and Holste, 2004). 
It can be explained as follows. As can be seen from 
Eq.(\ref{eq2}), $\Gamma_w(d)$ depends on both Cov() and 
Var(). While the denominator Var(GC\%) is dependent on
$w$ and gradually decreases with increasing window size,
the nominator Cov(GC\%) is practically independent of $w$. 
Experimentally, a slower-than-expected decrease of Var() 
with the window size was already observed around 1976 (Macaya et al., 1976). 
For random symbolic sequences, it can be shown by the binomial 
distribution that the variance of GC\% decreases with the window 
size $w$ according to Var(GC\%)$\sim 1/w$ 
(Nekrutenko and Li, 2000; Clay et al., 2001).
Sequence analysis shows,
however, that  Var(GC\%) $\sim 1/w^\beta$ with $0 < \beta < 1$
(Clay et al., 2001; Li and Holste, 2004).  This ``resistance to reduction of variance" is directly 
related to the $1/f^\alpha$ ($\alpha \approx 1$)
power spectrum (Clay et al., 2001; Clay et al., 2003; Clay, 2003; Li, 2005)
previously observed in DNA sequences (Li and Kaneko, 1992)  by
the relationship $\alpha \approx  1-\beta$.

Combining the lack of influence on Cov() by the window size,
and the factor of $1/w^\beta$ for  Var(), $\Gamma(d)$ is
expected to increase with increasing window size as $\Gamma_w(d)
\sim w^\beta$. Indeed, in Fig.2 $\log( \Gamma_w(d) )$ at $d=$1~Mb 
increases with $\log (w) $ more or less linearly. Excluding the
outlying chromosomes (15, 21, 22, X, Y), the regression
coefficient in Fig.2 is $\beta \approx 0.52$. This value
of $\beta$ is consistent with decay
exponents by the spectral analysis (Li and Holste, 2004), but note
that (i) it is an average among different chromosomes,
whereas the parameter fitting in (Li and Holste, 2004)  is carried
out separately on individual chromosomes; and (ii) a particular 
distance $d \approx$ 1~Mb is chosen.

Fig.2 clearly shows that considering the trend 
$\Gamma_w(d) \sim w^\beta$ (at $d=$1~Mb), chromosomes 15, 22, and Y 
have lower correlations than an average human chromosome, while 
chromosome 21 has higher correlations than the remaining chromosomes.
The lower correlation in chromosome Y is caused by the large
portion of unsequenced bases (about 50\%) and their substitution
with random values (Li and Holste, 2004). In the next
section, we will examine closely the causes of the unusual
high correlations in chromosome 21.

\section{An oscillation of 500~kb in the correlation function
is located at the rightmost one-eighth of the chromosome}

\indent

Fig.3 shows the correlation function $\Gamma(d)$ 
(for GC\% series calculated at the window size of $w=$358 bases)
as the function of $d$, 
for the DNA sequences of human chromosome 21. There is a striking 
oscillation in $\Gamma(d)$ which peaks at about 0.5, 1.1, 1.6,
and 2.1~Mb with an approximately constant spacing between peaks of 
about 500~kb.  From Fig.1, it can be seen this long-range 
oscillation is solely present in human chromosome 21, but not in 
other human chromosomes. As discussed above, a sequence periodicity 
can be caused either by an exact repeat or by a tendency
for a particular base to be located in a periodic location. 
In either case, to maintain base-level periodicity for such
a distance of hundreds of kb or longer requires a
selection pressure against insertion and deletion mutations.

To find out whether this 500~kb periodicity is localized
in a particular region on chromosome 21, we segment
chromosome 21 evenly into eight segments. Fig.4 shows the 
correlation function $\Gamma(d)$ for each segment  
as a function of $d$. The first and the second segments 
are mostly unsequenced, and hence $\Gamma(d)$ is flat
as unsequenced bases are substituted by 
random bases.  In the next three chromosomal regions 
(11.7~Mb- 29.3~Mb), essentially no apparent correlation structure 
is present in $\Gamma(d)$ at $d=$ 500~kb $\sim$ 2.5~Mb range.
The gradual decay of $\Gamma(d)$ from 500~kb to 1.5~Mb
in the sixth chromosomal segment (29.3~Mb-35.2~Mb) is mainly due to
an onset from an L2-isochore to an H1-isochore 
(Bernardi, 2001; Pavl\'{i}\u{c}ek et al., 2001).
The correlation function  of the last segment (41.1~Mb-46.9~Mb)
reveals the source of the 500~kb oscillation: the peak 
locations are exactly the same as those in Fig.3 albeit
without the decay trend. This segment corresponds largely to 
the GC-rich isochore of chromosome 21. 

Fig.5 shows the chromosomal region, from 43.5~Mb to 46.5~Mb, 
that exhibits the 500~kb oscillation. Fig.5(a) shows the 
GC\% calculated from the window size of $w=$ 2,864 bases. 
Also shown in Fig.5(a) is a sinusoidal function with the
period of 500~kb. The second half of the sinusodial
function is shifted from the first half to better fit 
the oscillation in the GC\% fluctuation. It can be seen that six 
GC-high peaks alternate with five GC-low
valleys, though the third valley is not as low 
as the others.  The distance between two GC-high peaks 
(or GC-low valleys) is approximately equal to 500~kb, with the exception 
of the middle region that has a longer spacing between 
peaks. It is interesting to
note that the alteration of GC-high and GC-low regions,
but not the regularity of the spacing, had implicitly been 
detected before (e.g. Fig.1 of Hattori et al., 2000, 
and Fig.3 of Bernardi, 2001).

Because interspersed repeats tend to have higher GC\%
than the rest of the sequence, we address the question
of whether a regular spatial distribution of the repeat
sequences is responsible for the observed 500~kb oscillation.
Fig.5(b) shows a similar GC\% fluctuation for the 
sequence with substituted interspersed repeats.  The 500~kb oscillation 
persists even when interspersed repeats are substituted.

\section{Discussion}

\indent

In this paper, we have observed a localized 500~kb long 
oscillation in GC\% of human chromosome 21. We checked
whether the region of chromosome 21 with this oscillation 
has been the focus of investigations in previously 
large-scale correlation analyses of the human genome,
and we found that a segmental duplication of size 
of 200~kb has been identified on chromosome 21 
(Golfier et al., 2003).  However, the region reported in
(Golfier et al., 2003)  was in the chromosome band 21q22.1, 
whereas the last one-eighth segment of the human chromosome 21 
reported here was in the band 21q22.3.

The 21q22.3 band is both GC-rich and gene-rich, in marked contrast 
to the 7~Mb GC-poor isochore localized in 21q21.1-21q21.2
(Hattori et al., 2000).
There are 68 known genes within the position of 43.5 -- 46.5~Mb,
or roughly one gene per 44~kb. As a comparison, there are
total 268 genes for the whole chromosome 21 with length of 46.9~Mb,
or one gene per 175~kb. Some of the genes are of interests
to human disease gene mapping. For example, a rare autoimmune 
disease that affects the endocrine glands, called autoimmune 
polyglandular syndrome type I (APS1) or autoimmune 
polyendocrinopathy-candidiasis-ectodermal dystrophy (APECED) 
(Online Mendelian Inheritance in Man (McKusick, 1998) number 240300), 
was shown to be linked to markers in 21q22.3 
(Aaltonen et al., 1994).  The linked region is further narrowed down to 
the autoimmune regulator (AIRE) gene 
(Aaltonen et al., 1997; Nagamine et al., 1997),
located at the position 44.561-44.574~Mb.

Fig.5(c) shows the location of mapped known genes,
using the {\sl knownGene} field of the UCSC genome bioinformatics
site ({\sl http://genome.ucsc.edu/goldenPath/gbdDescriptions.html}).
Genes on two DNA strands are plotted separately. When 
known genes on both strands are
considered together, there is no visible gaps in their
spatial distribution. However, strand-specific gene
distribution seems to match the amplitudes of GC\% oscillations.
We observe that the forward and the reverse strand alternately 
exhibits comparatively lower local gene densities 
in GC-poor valleys:  1 and 3 for  the reverse strand ($-$),
2 and 4 for the forward strand ($+$). Both strands show
a lower local gene density in GC-poor valley region 5.

We next investigated the spatial distribution of quantities 
that are directly related to GC\%. Fig.5(d) shows 
``long homogeneous genome regions" (Oliver et al., 2001) 
as detected by the program {\sl IsoFinder} (Oliver et al., 2004).
There are two interesting immediate observations: Firstly, the
third valley is not as GC-poor as predicted by the
sinusoidal function, which can also be confirmed by examining 
Fig.5(a) and (b).  Secondly, there is a lack of periodic alternation
between the GC-rich and GC-poor segments with comparable 
sizes. This can be explained by the difference in the
segment or isochore view of global GC\% variation. Isochores 
corresponds to GC\% fluctuation that can be approximated
by step functions. Gradual changes of GC\% as captured by 
sinusoidal functions are approximated poorly by step functions.

Fig.5(d) also shows the location of CpG islands (map extracted
from the UCSC genome bioinformatics site). A visual inspection
shows that CpG islands are rare in GC-poor valleys, in particular
valleys No. 1, 3, 4, and 5. It is not a completely unexpected
observation since one of the criteria for CpG island detection
is its GC\%. In particular, one of the oftenly used methods for 
CpG island detection requires GC\% to be higher than 50\% 
(Larsen et al., 1992).
The GC-high peaks and GC-poor valleys in Fig.5 are separated
by the GC\%=50\% line, and this may explain why CpG
islands are less likely to be found in these GC-poor valleys.

In summary, we have detected a unique long-range oscillation 
in a localized region in human chromosome 21, which is 
absent in the remaining human chromosomes.  It will be of 
interests to determine the key biological features either 
causing or resulting from this ultralong-ranging periodicity 
in human chromosome 21.  While it cannot be excluded that 
this particular oscillation is due to chance events, 
one of the promising directions to pursue is
its connection to chromatin structure and DNA loops,
along the similar line of research on the connection
between these structural units and GC\% 
(Saccone et al., 2002; Bernardi, 2004).

\section*{Acknowledgements}

\indent

We would like to thank J\'{o}se Oliver, Pedro Miramontes, 
Chun-Ting Zhang, Peter Gregersen, Atsushi Fukushima for
discussions, Oliver Clay, Pedro Bernaola-Galv\'{a}n 
for a critical reading of the manuscript, and the UCSC
bioinformatics group for providing genome resources 
to the community.

\renewcommand{\baselinestretch}{1.1} % double spacing

\section*{References}

\vspace{0.09in}
\noindent
Aaltonen, J., Bj\"{o}rses, P., Sandkuijl, L., Perheentupa, J., Peltonen, L., 1994.
An autosomal locus causing autoimmune disease, autoimmune polyglandular 
disease type I assigned to chromosome 21.
Nature Genetics, 8,83-87. 

\vspace{0.09in}
\noindent
Aaltonen, J., {\sl et al.}  (The Finnish-German APECED consortium), 1997.
An autoimmune disease, APECED, caused by mutations in novel gene featuring two
PHD-type zinc-finger domains.
Nature Genetics, 17,339-403.

\vspace{0.09in}
\noindent
Arques, D.G. 1987.
Periodicities in introns.
Nucleic Acids Research, 15, 7581-7592.

\vspace{0.09in}
\noindent
Baldi, P., Brunak, S., Chauvin, Y., Krogh, A., 1996. 
Naturally occurring nucleosome positioning signals in 
human exons and introns. 
Journal of Molecular Biology, 263,503-510. 

\vspace{0.09in}
\noindent
Bernardi, G , 2001.
Misunderstandings about isochores. Part 1.
Gene, 276,3-13.

\vspace{0.09in}
\noindent
Bernardi, G , 2004.
{\sl Structural and Evolutionary Genomics}
(Elsevier).

\vspace{0.09in}
\noindent
Borodovsky, M., McIninch, J., 1993.
GeneMark, parallel gene recognition for both DNA strands.
Computers \& Chemistry, 17,123-133.
 
\vspace{0.09in}
\noindent
Borodovsky, M., Sprizhitskii, Yu., Golovanov, E., Aleksandrov, A., 1986.
Statistical patterns in primary structures of functional regions 
in the E. coli genome. I. oligonucleotide frequencies analysis.
Molecular Biology, 20,826-833;
II. Non-homogeneous Markov models. {\sl ibid},  20,833-840;
III Computer recognition of coding regions. {\sl ibid}, 20,1145-1150.

\vspace{0.09in}
\noindent
Burge, C., Karlin, S., 1997.
Prediction of complete gene structures in human genomic DNA.
Journal of Molecular Biology, 268,78-94.

\vspace{0.09in}
\noindent
Clay, O , 2001.
Standard deviations and correlations of GC levels in DNA
sequences.
Gene, 276,33-38.

\vspace{0.09in}
\noindent
Clay, O., Carels, N., Douady, C., Macaya, G., Bernardi, G., 2001.
Compositional heterogeneity within and among isochores
in mammalian genomes.
Gene, 276,15-24.

\vspace{0.09in}
\noindent
Clay, O., Douady, C.J., Carels, N., Hughes, S., Bucciarelli, G., Bernardi, G., 2003.
Using analytical ultracentrifugation to study compositional
variation in vertebrate genomes.
European Biophysics Journal, 32,418-426.

\vspace{0.09in}
\noindent
Fickett, J.W., 1982.
Recognition of protein coding regions in DNA sequence.
Nucleic Acids Research, 10,5303-5318.

\vspace{0.09in}
\noindent
Fukushima, A., Ikemura, T., Kinouchi, M., Oshima, T., Kudo, Y., Mori, H., 
Kanaya, S., 2002. 
Periodicity in prokaryotic and eukaryotic genomes identified by power 
spectrum analysis.
Gene, 300,203-211. 

\vspace{0.09in}
\noindent
Golfier, G., Chibon, F., Aurias, A., Chen, X.N., Korenberg, J., Rossier, J., 
Potier, M.C., 2003.
The 200-kb segmental duplication on human chromosome 21 originates
from a pericentromeric dissemination involving human chromosomes 2, 18 
and 13.
Gene, 312,51-59.

\vspace{0.09in}
\noindent
Grosse, I., Herzel, H., Buldyrev, S.V., Stanley, H.E., 2000. 
Species independence of mutual information in coding 
and noncoding DNA.
Physical Review, 61, 5624-5629. 

\vspace{0.09in}
\noindent
Guigo, R., 1999.
DNA composition, codon usage and exon prediction.
in {\sl Genetics Databases}, ed. M Bishop , Academic Press.

\vspace{0.09in}
\noindent
Hattori, M., {\sl et al.} , 2000.
The DNA sequence of human chromosome 21.
Nature, 405,311-319.

\vspace{0.09in}
\noindent
Herzel, H., Weiss, O., Trifonov, E.N., 1998. 
Sequence periodicity in complete genomes of Archaea suggests 
positive supercoiling.
Journal of Biomolecular Structure and Dynamics, 16,341-345. 

\vspace{0.09in}
\noindent
Holste, D., Grosse, I., Beirer, S., Schieg, P.,  Herzel, H., 2003.
Repeats and correlations in human DNA sequences.
Physical Review E, 67,061913.

\vspace{0.09in}
\noindent
Konopka, A.K., Smythers, G.W., Owens, J., Maizel, J.V. Jr.  1987.
Distance analysis helps to establish
characteristic motifs in intron sequences.
Gene Analysis Techniques, 4, 63-74.

\vspace{0.09in}
\noindent
Larsen, F., Gundersen, G., Lopez, R., Prydz, H., 1992.
CpG islands as gene markers in the human genome.
Genomics, 13,1095-1107.

\vspace{0.09in}
\noindent
Li, W., 2005.
Large-scale fluctuation of guanine and cytosine content
in genome sequences, isochores and 1/f spectra.
in {\em Progress in Bioinformatics}. Nova Science,
Hauppauge, NY.

\vspace{0.09in}
\noindent
Li, W., Bernaola-Galv\'{a}n, P., Haghighi, F., Grosse, I., 2002. 
Applications of recursive segmentation to the analysis 
of DNA sequences.
Computers \& Chemistry, 26, 491-510. 

\vspace{0.09in}
\noindent
Li, W., Holste, D., 2004.
Universal 1/f noise, cross-overs of scaling exponents, 
and chromosome specific patterns of GC content in
DNA sequences of the human genome.
submitted to Physical Review E.

\vspace{0.09in}
\noindent
Li, W., Kaneko, K., 1992.
Long-range correlation and partial $1/f^{\alpha}$ spectrum in a
non-coding DNA sequence.
Europhysics Letters, 17,655-660.

\vspace{0.09in}
\noindent
Macaya, G., Thiery, J.P., Bernardi, G., 1976.
An approach to the organization of eukaryotic genomes
at a macromolecular level.
Journal of Molecular Biology, 108,237-254.

\vspace{0.09in}
\noindent
McKusick, V.A. , 1998.
{\sl Mendelian Inheritance in Man. A Catalog of Human Genes and Genetic Disorders}, 
Johns Hopkins University Press, Baltimore. 12th edition. 
URL, {\sl http,//www.ncbi.nlm.nih.gov/omim/}.

\vspace{0.09in}
\noindent
Nagamine, K., {\sl et al.} , 1997.
Positional cloning of the APECED gene.
Nature Genetics, 17,393-398. 

\vspace{0.09in}
\noindent
Nekrutenko, A., Li, W.H. , 2000.
Assessment of compositional heterogeneity within and between
eukaryotic genomes.
Genome Research, 10,1986-1995.

\vspace{0.09in}
\noindent
Oliver, J.L., Bernaola-Galv\'{a}n, P.,  Carpena, P., Rom\'{a}n-Rold\'{a}n, R., 2001.
Isochore chromosome maps of eukaryotic genomes.
Gene, 276,47-56.

\vspace{0.09in}
\noindent
Oliver, J.L., Carpena, P., Hackenberg, M., Bernaola-Galv\'{a}n, P., 2004.
IsoFinder, computational prediction of isochores in genome sequences.
Nucleic Acids Research,  32,W287-W292.

\vspace{0.09in}
\noindent
Pavl\'{i}\u{c}ek, A., et al. , 2001.
Similar integration but different stability of Alus and LINEs in the
human genome.
 Gene, 276,39-45.

\vspace{0.09in}
\noindent
Pizzi, E., Liuni, S., Frontali, C. , 1990.
Detection of latent sequence periodicities.
 Nucleic Acids Research, 18,3745-3752.

\vspace{0.09in}
\noindent
Saccone, S., Frederico, C., Bernardi, G. 2002.
Localization of the gene-richest and the gene-poorest isochores 
in the interphase nucleic of mammals and birds.
Gene, 300, 169-178. 

\vspace{0.09in}
\noindent
Shiba, K., Takahashi, Y., Noda, T., 2002.
On the role of periodism in the origin of proteins.
Journal of Molecular Biology, 320, 833-840.

\vspace{0.09in}
\noindent
Smit, A.F. , 1999. 
Interspersed repeats and other mementos of 
transposable elements in mammalian genomes.
Current Opinion in Genetics \& Development, 9, 657-663. 

\vspace{0.09in}
\noindent
Staden, R., McLachlan, A.D. , 1982.
Codon preference and its use in identifying protein coding regions 
in long DNA sequences.
Nucleic Acids Research, 10, 141-156.

\vspace{0.09in}
\noindent
Shulman, M.J., Steinberg, C.M., Westmoreland, N., 1981.
The coding  function of  nucleotide  sequences  can  be  discerned by 
statistical analysis.
Journal of Theoretical  Biology,  88, 409-420.

\vspace{0.09in}
\noindent
Tiwari, S., Ramachandran, S., Bhattacharya, A., Bhattacharya, S., Ramaswamy, R. , 
1997. Prediction of probable genes by Fourier analysis of 
genomic sequences.
Computer Applications in Biosciences, 13, 263-270. 

\vspace{0.09in}
\noindent
Tomita, M., Wada, M.,  Kawashima, Y., 1999. 
ApA dinucleotide periodicity in prokaryote, eukaryote, and organelle genomes.
Journal of Molecular Evolution, 49,182-192. 

\vspace{0.09in}
\noindent
Trifonov, E.N., Sussman, J.L., 1980. 
The pitch of chromatin DNA is reflected in its nucleotide sequence.
Proceedings of the National Academy of Sciences, 77,3816-3820. 

\vspace{0.09in}
\noindent
Trifonov, E.N., 1998.
3- 10.5- 200- and 400-base periodicities in
genome sequences.
Physica A, 249,511-516.

\vspace{0.09in}
\noindent
Widom, J., 1996. Short-range order in two eukaryotic genomes, 
relation to chromosome structure.
Journal of Molecular Biology, 259,579-588.

\vspace{0.09in}
\noindent
Zhurkin, V.B., 1981.
Periodicity in DNA primary structure is defined by 
secondary structure of the coded protein.
Nucleic Acids Research, 9,1963-1971.

\newpage

\large

\begin{figure}
\begin{center}
  \begin{turn}{-90}
  \epsfig{file=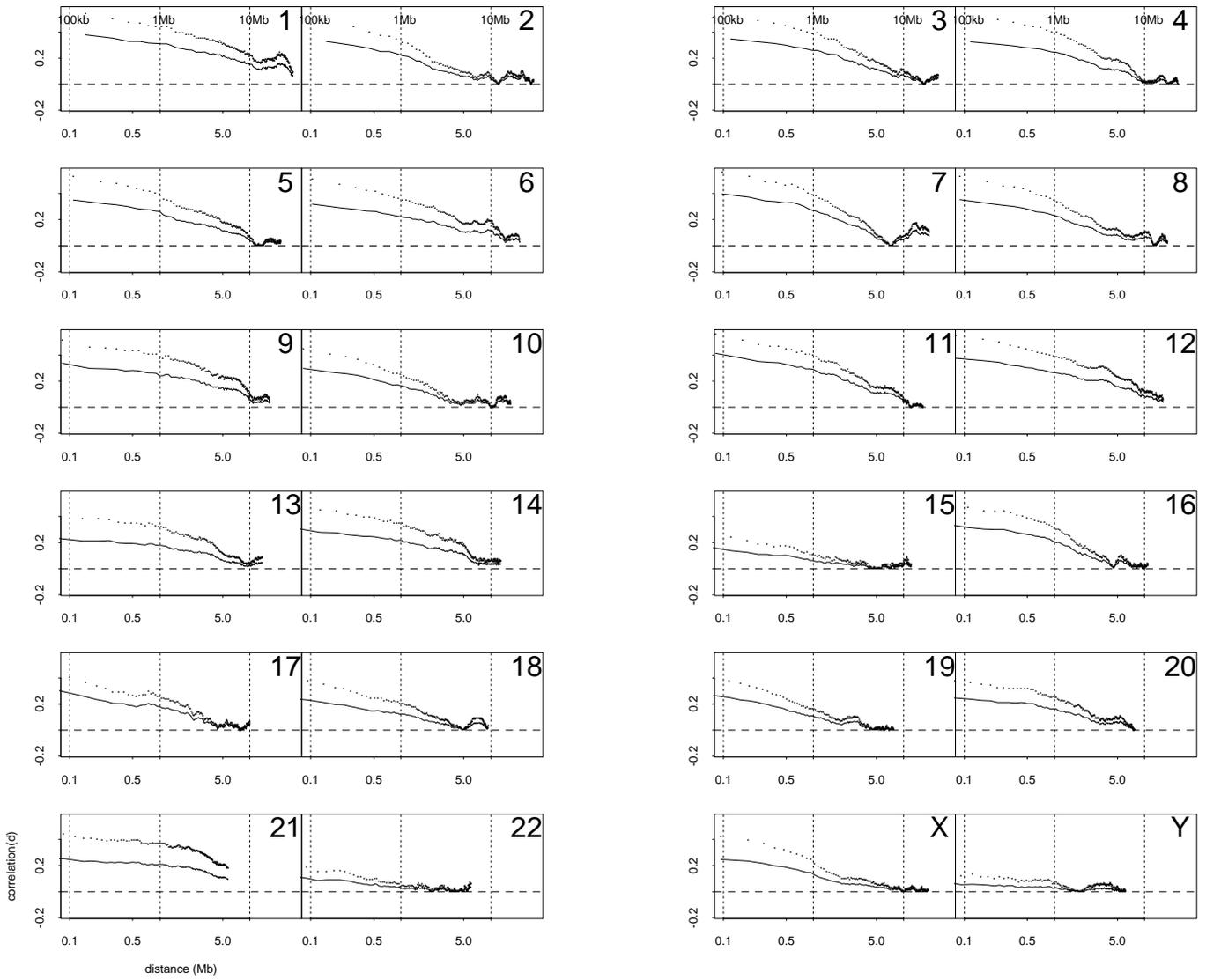, height=18cm}
  \end{turn}
\end{center}
\caption{
\label{fig1}
Correlation function $\Gamma(d)$ of the GC\% series with
$2^k$ GC values ($k$=15: dots, top;  and $k$=17: lines, 
bottom), obtained from all 24 human chromosomes 
(22 autosomal and 2 sex chromosomes).
The x-axis, in a logarithmic scale, is the distance $d$ converted 
to units of Mb ($10^6$ bases). 
}
\end{figure}

\begin{figure}
\begin{center}
  \begin{turn}{-90}
  \epsfig{file=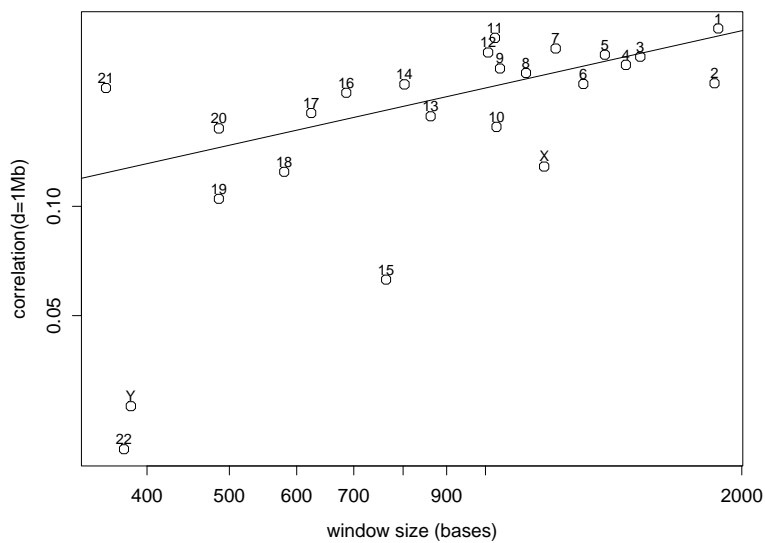, height=10cm}
  \end{turn}
\end{center}
\caption{
\label{fig2}
Correlation $\Gamma_w(d)$,  at the distance $d \approx $ 1~Mb, 
as a function of the chromosome-specific  window size $w$
(in a double logarithmic plot).  Each point represents one human 
chromosome. The window size $w$ is the chromosome length 
divided by $2^{17}$.
}
\end{figure}

\begin{figure}
\begin{center}
  \begin{turn}{-90}
  \epsfig{file=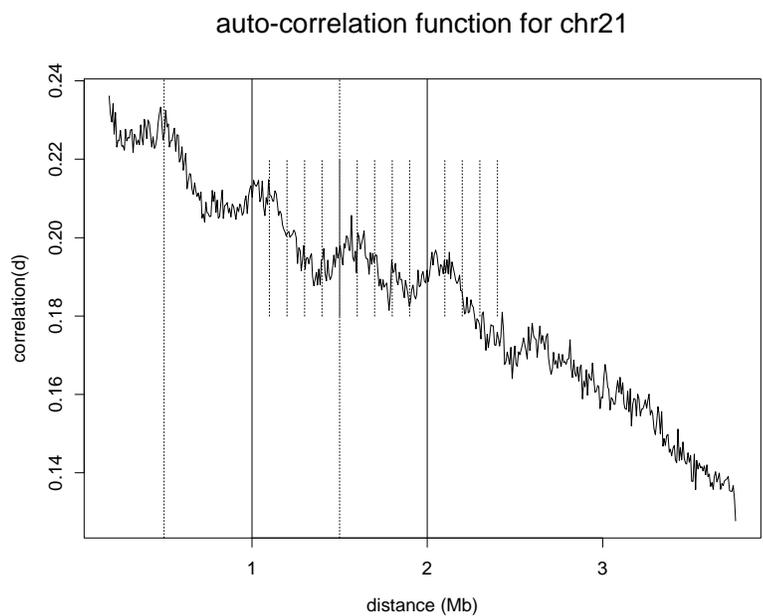, height=10cm}
  \end{turn}
\end{center}
\caption{
\label{fig3}
Correlation function $\Gamma(d)$ of the GC\% series for the
DNA sequence of human chromosome 21.
GC\% is calculated at the window size of 358 bases, which is
$1/2^{17}$ of the total chromosome length. The
distances of 0.5, 1, 1.5, and 2~Mb are marked by long vertical lines,
and the spacing of 100~kb is marked by short vertical lines.
}
\end{figure}

\begin{figure}
\begin{center}
  \begin{turn}{-90}
  \epsfig{file=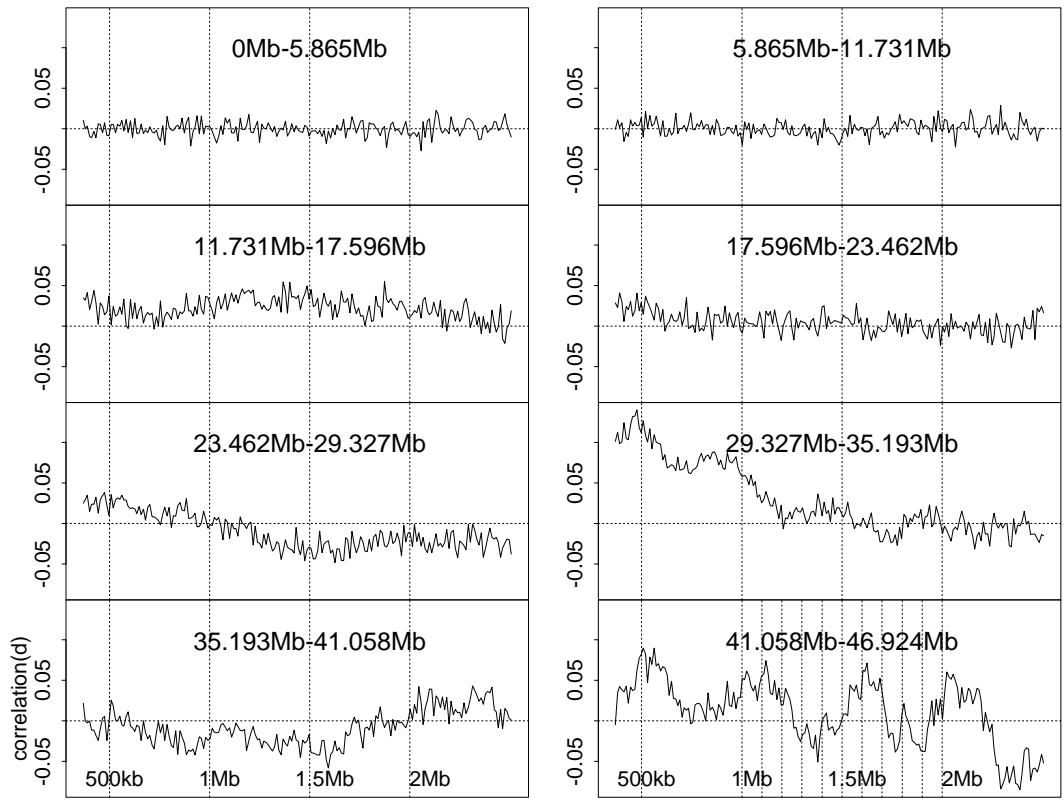, height=14cm}
  \end{turn}
\end{center}
\caption{
\label{fig4}
Correlation function $\Gamma(d)$ of the GC\% series for
eight chromosomal regions of the DNA sequence
of the human chromosome 21. The distances of
0.5, 1, 1.5, and 2~Mb are marked by vertical lines.
}
\end{figure}

\begin{figure}
\begin{center}
  \begin{turn}{-90}
  \epsfig{file=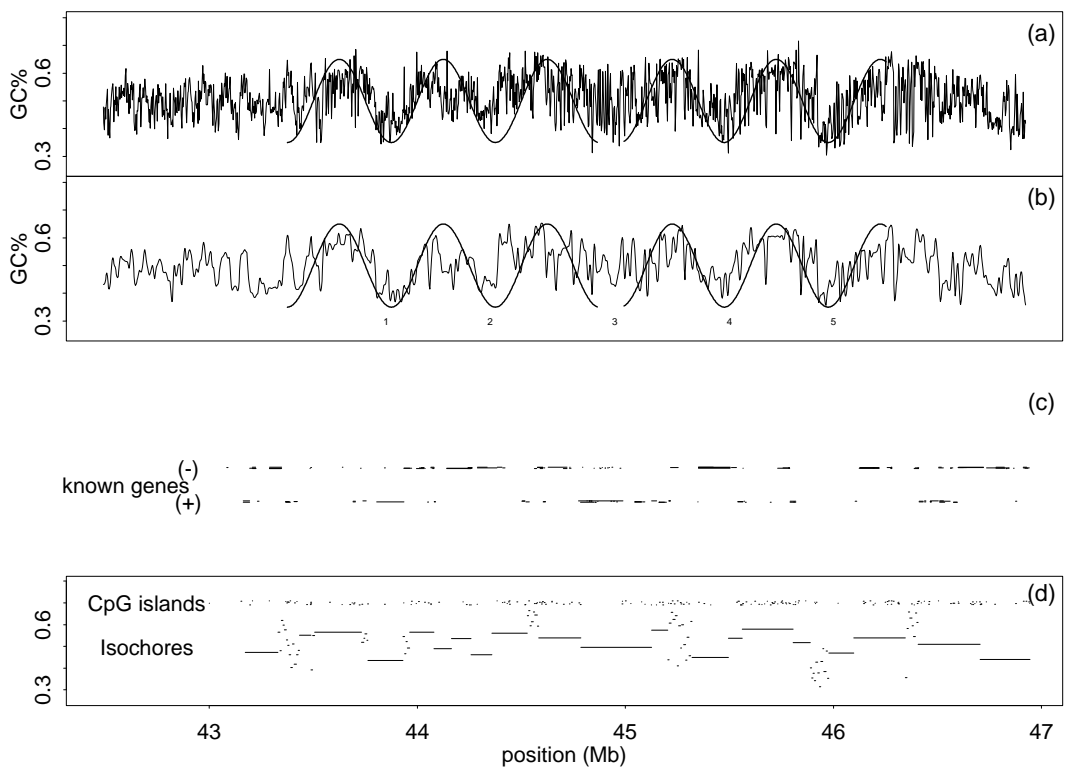, height=14cm}
  \end{turn}
\end{center}
\caption{
\label{fig5}
GC\% fluctuations of the last one-eighth segment of the 
DNA sequence of human chromosome 21
towards the q-term end. (a) GC\% calculated
for the window size of $w=$ 2.864~kb, which is $1/2^{14}$
of the whole chromosome length. A sinusoidal function with 
the period of 500 kb is superimposed on the plot to fit 
the periodic oscillation of GC\%.  (b) GC\% calculated with 
interspersed repeats replaced by random values, then smoothed 
by means of running medians, using the {\sl S-PLUS} subroutine 
{\sl smooth }.
(c) 
Locations of known genes as determined by protein
sequences from SWISS-PROT, TrEMBL, and TrEMBL-NEW, and 
their corresponding mRNAs from GenBank, displayed separately 
for each DNA strand. 
(d) Locations of CpG islands (top) and isochores (bottom). 
For the isochore map,  the GC\% of individual isochores
is indicated by the height of the horizontal bar. 
}
\end{figure}

\end{document}